\documentclass[aip,pop,reprint,numerical]{revtex4-1}
\usepackage[utf8]{inputenc}

\usepackage{amssymb}
\usepackage{amsfonts,amstext} 
\usepackage{amsmath}
\usepackage{bm}
\usepackage{multirow}
\usepackage{graphicx}
\usepackage[space]{grffile}
\usepackage{xspace}
\usepackage{color}
\usepackage[usenames,dvipsnames]{xcolor}  
\usepackage[font=small,labelfont=bf]{subcaption}

\usepackage[normalem]{ulem}
\usepackage{wasysym}
\usepackage{aas_macros}
\usepackage{gensymb}
\usepackage{booktabs}

\usepackage[hidelinks]{hyperref}  
\hypersetup{colorlinks=false}
\usepackage{url}

\usepackage{xparse}
\usepackage{soul}
\usepackage{my_commands}
\usepackage{cleveref}

\newcommand{\REV}[1]{#1}

\makeatletter
\def\widebreve#1{\mathop{\vbox{\m@th\align{##\crcr\noalign{\kern3\p@}%
				\brevefill\crcr\noalign{\kern3\p@\nointerlineskip}%
				$\hfil\displaystyle{#1}\hfil$\crcr}}}\limits}

\def\brevefill{$\m@th \setbox\z@\hbox{$\braceld$}%
	\bracelu\leaders\vrule \@height\ht\z@ \@depth\z@\hfill\braceru$}

\begin{document}

		\title{\REV{A nonlinear structural subgrid-scale closure for compressible MHD Part I: derivation and energy dissipation properties}}
		\author{Dimitar G Vlaykov}
		\email{Dimitar.Vlaykov@ds.mpg.de}
		\affiliation{Institut für Astrophysik, Universität Göttingen, Friedrich-Hund-Platz 1, D-37077 Göttingen, Germany}
		\affiliation{Max-Planck-Institut für Dynamik und Selbstorganisation, Am Faßberg 17, D-37077 Göttingen, Germany}
		\author{Philipp Grete}
		\affiliation{Max-Planck-Institut für Sonnensystemforschung, Justus-von-Liebig-Weg 3, D-37077 Göttingen, Germany}
		\affiliation{Institut für Astrophysik, Universität Göttingen, Friedrich-Hund-Platz 1, D-37077 Göttingen, Germany}
		\author{Wolfram Schmidt}
		\affiliation{Hamburger Sternwarte, Universität Hamburg, Gojenbergsweg 112, D-21029 Hamburg, Germany}
		\author{Dominik R G Schleicher}
		\affiliation{Departamento de Astronomía, Facultad Ciencias Físicas y Matemáticas, Universidad de Concepción,  Av. Esteban Iturra s/n Barrio Universitario, Casilla 160-C, Chile}
		
		\begin{abstract}
		Compressible magnetohydrodynamic (MHD) turbulence is ubiquitous in astrophysical phenomena ranging from the intergalactic to the stellar scales. 
		In studying them, numerical simulations are nearly inescapable, due to the large degree of nonlinearity involved.
		However the dynamical ranges of these phenomena are much larger than what is computationally accessible.
		In large eddy simulations (LES), the resulting limited resolution effects are addressed explicitly by 
		introducing to the equations of motion additional terms associated with the unresolved, subgrid-scale (SGS) dynamics.
		This renders the system unclosed.
		We derive a set of nonlinear structural closures for the ideal MHD LES equations with particular emphasis on the effects of compressibility. 
		The closures are based on a gradient expansion of the finite-resolution operator\cite{Yeo1993} and require no assumptions about the nature of the flow or magnetic field. Thus the scope of their applicability ranges from the sub- to the hyper-sonic  and -Alfvenic regimes.
		The closures support spectral energy cascades both up and down-scale, as well as direct transfer between kinetic and magnetic resolved and unresolved energy budgets. They implicitly take into account the local geometry, and in particular the anisotropy, of the flow. 
		Their properties are \aprio validated in an accompanying article\cite{Grete2016} against alternative closures available in the literature with respect
		to a wide range of simulation data of homogeneous and isotropic turbulence.

		\end{abstract}
	\noindent {\textit{Compressible Magnetohydrodynamics, Scale separation, Subgrid-scale closure, Turbulence}}
	\date{\today}
	
	\maketitle			
\section{Introduction}

There is a great need for increased accuracy in numerical simulations involving turbulent flows of magnetized fluids in fields varying from engineering to astrophysics. In astrophysics, in particular, compressible magnetohydrodynamic (MHD) turbulence is an important ingredient in the solution of outstanding problems on many scales such as the generation and sustainment of galactic and super-galactic scale magnetic fields\cite{Subramanian2006, Cho2014, Latif2013}; the detailed process of star formation, including  self-regulation  and fragmentation\cite{Low2004, Price2008, Hennebelle2008}; stellar convection in the interior and stellar atmospheres \cite{Balbus1998};  accretion and protoplanetary discs, stellar ejecta, e.g. jets, winds, outflows\cite{Biskamp2003, Pudritz2007}; the dynamics of the solar tachocline, the solar wind and the solar corona 
\cite{Priest2014, Zhou2004, Mangeney1991, Chernyshov2010, Bruno2013}. The dynamical range of these phenomena is usually much larger than what is computationally tractable. Numerically, this translates to unphysical dissipation and turbulence dynamics due to the limited resolution. For example, in finite-volume numerical schemes \REV{it} leads to enhanced dissipation. In large eddy simulations (LES)\cite{Sagaut2006, Sagaut2009, Miesch2015, Schmidt2015} this problem is tackled by directly solving only the evolution equations for the resolved fields. The \REV{contribution} of the small under- and unresolved scales (i.e. the scales which are badly contaminated by numerical noise or simply unrepresented) on them has to be incorporated \REV{via explicit modeling}. 
Formally these  scales are identified by the introduction of a finite resolution operator, in effect a low-pass filter. Large eddy simulations are typically used with grid-based numerical schemes, e.g. \REV{based on finite-differences or finite-volumes}. As such the grid-scale can be taken to be the filter scale and hence the terms responsible for the small-scale effects are known as subgrid-scale (SGS) terms. 

The magnetohydrodynamic LES  equations are obtained by applying a finite resolution operator to the MHD equations. It can be shown that this operator can be expressed as a convolution with a low-pass filter kernel. There are several comprehensive reviews of the formalism and its
application to hydrodynamics\cite{Sagaut2006, Sagaut2009, Schmidt2015} and MHD \cite{Chernyshov2014}.
Applying the formalism with a static, homogeneous and isotropic kernel $G$ with a constant grid-scale (which can be used to represent the commonly used grid-based numerical schemes in physical or spectral space) under periodic boundary conditions to the ideal MHD equations results in the following equations for the large-scale fields:
\begin{subequations}
\label{eq: les}
\begin{align}
\label{eq: les mass}
\pd{\rres}{t}+\div {\rres \fav{\v{u}}} = 0,\\
\label{eq: les mom}
\pd{\flt{\rho} \fav{\v{u}}}{t}
+ \div{\flt{\rho} \fav{\v{u}} \otimes \fav{\v{u}}
	- \flt{\v{B}} \otimes \flt{\v{B}}}
+ \grad{\pres + \frac{\flt{B}^2}{2}} =&- \nabla \cdot \tau,\\ 
\label{eq: les ind}
\pd{\flt{\v{B}}}{t} - \curl\bra{\fav{\v{u}} \times \flt{\v{B}}} =
\curl \emf-.
\end{align}
\end{subequations}
Here a large scale, filtered field is denoted by an overbar. For instance, the large scale component of the pressure $P$ is given by a convolution with the filter kernel $G$, i.e.   $\filt{P}=G\ast P$ and similarly for the filtered density $\rres$  and the magnetic field  $\bres-$.  The treatment of the pressure term is beyond the scope of this work due to the wide array of possible equations of state used to close the MHD system. Nevertheless, briefly, if the equation of state is linear in the primary fields (e.g. in isothermal conditions), the pressure does not lead to any SGS contributions.\REV{\newline}
The tilde denotes a mass-weighted (also known as Favre) filtered field\cite{Favre1983}, i.e.~the Favre-filtered velocity field $\ures=\flt{\rho\v{u}}/\rres$.
\REV{Using $\ures$ as a primary quantity} precludes the introduction of SGS terms in the mass conservation equation\REV{}. Additionally, it fits well with physical-space-based compressible schemes, where often the momentum $\rho \v{u}$ is evolved as the primary quantity instead of the velocity $\v{u}$.
The momentum and induction equations contain two new, SGS terms, $\div \tau$ and $\curl \emf-$, which will occupy the focus of this article. They are simply the commutators between the finite resolution operator and the nonlinearities of the respective MHD equations. \REV{Thus} they carry information about the interactions across the filter scale.\REV{} Analytically they are given by
\begin{align}
  \label{eq:tau_emf_def}
  \nonumber \emf- &= \flt{\v{u}\times\v{B}} - \ures- \times\bres- ,\; \mathrm{and}\\\nonumber  
\tau[][ij] &= \tau[u][ij]-\tau[b][ij]+\half\tau[b][kk]\dirac\;\;\mathrm{ with}, \\
\tau[u][ij]&= \flt{\rho} \bra{\fav{u_i u_j} - \fav{u}_i \fav{u}_j},\;\; \tau[b][ij]= \bra{\flt{B_i B_j} - \flt{B}_i~\flt{B}_j},
\end{align} 
where the Einstein summation convention is assumed.
The tensor $\tau$ is known as the SGS stress and can be decomposed into kinetic and magnetic components, SGS Reynolds stress $\tau[u]$ and SGS Maxwell stress $\tau[b]$ respectively. The (pseudo-)vector $\emf-$ is known as the electromotive force. 
\REV{They carry information about the subgrid-scales via the terms $\flt{\v{u}\times\v{B}}$, $\fav{u_i u_j}$, and $\flt{B_i B_j}$ and thus cannot be explicitly expressed only in terms of large scale fields. This renders} the
system of equations (\ref{eq: les}) unclosed. The evolution equations of the SGS terms\cite{Sagaut2006}, involve new, higher order unknown terms.
This continues to build an infinite hierarchy. This is the LES aspect of the well-known turbulence closure problem. 
\REV{\newline}
\REV{The resolved, i.e large scale, energies and cross-helicity} are defined as
\begin{align}
&\ekinres = \half \rres \ures-^2, \; \emagres = \half \bres-^2,\;\eres = \ekinres+\emagres\, ,\\ \nonumber
& \textrm{and }W_\res = \ures-\cdot \bres-.\;
\end{align}
Their evolution equations are obtained in the classical manner from the corresponding
primary LES equations\cite{Vlaykov2015}. For ideal MHD they can be written as
\begin{align}
\label{eq: kinenres}\nonumber
\pd{}{t}\ekinres +
\div{\ures-\ekinres}
+\ures- \cdot \bres- \times \jres-
+ \ures- \cdot\grad \pres  &=\\- \ures-\cdot\bra{\nabla \cdot
	\tau},\\
\label{eq: magenres}\nonumber
\pd{}{t}\emagres - \bres-\cdot\curl\bra{\ures- \times
	\bres-} &=\\
\bres- \cdot \curl \emf-,\\
\label{eq: totenres}\nonumber
\pd{\eres}{t} +
\div{\ures-\ekinres+
	2\ures- \emagres-
	\bres- W_\res}+ \ures- \cdot \grad \flt{P} &=\\ \bres- \cdot \curl \emf- - \ures-\cdot\bra{\nabla \cdot
	\tau} ,\\
\label{eq: crhelres}\nonumber
\pd{}{t}W_\res+ \div{ \ures- W_\res -  \frac{\bres-}{\rres} \ekinres}
+   \frac{\bres-}{\rres}\cdot \grad \pres &= \\
\ures- \cdot \curl\emf- - \frac{\bres-}{\rres}
\cdot \bra{\div \tau},
\end{align}
where $\jres-=\nabla \times \bres-$ is the resolved current.
Although the total energy and cross-helicity are ideal MHD invariants, their resolved counterparts, as defined above, are not, 
due to the SGS terms on the right hand side of \cref{eq: totenres,eq: crhelres}. \REV{The equations show that} the SGS stress and EMF encode the entire transfer of energy and cross-helicity across the filter scale \REV{and truncating the SGS hierarchy at the level of $\tau$ and $\emf-$ closes these equations as well}.
 
Various approaches have been developed to address the closure problem
for hydrodynamics\cite{Sagaut2009, Sagaut2006}, in astrophysical settings\cite{Schmidt2015}. Several models have also been extended to the case of magnetized fluids\cite{Muller2002, Muller2002a, Pietarila2009}, some of them taking into account compressibility as well\cite{Chernyshov2014, Grete2015}.
They can be separated heuristically into structural and functional ones.
\REV{Functional closures focus on the effect of the SGS terms on the resolved scales and are thus largely phenomenological. For instance, the eddy-viscosity models\cite{Chernyshov2014} address the anomalous energy dissipation due to turbulence, while dynamo models\cite{Widmer2015,Yokoi2013} address the generation and amplification of magnetic fields.
Structural models try to mimic some aspect of the structure of the SGS terms, expecting that the desired effects on the large scale will follow automatically. Thus they largely rely on the robustness of these aspects. In the self-similarity closures\cite{Bardina1983,Chernyshov2014} for example, the main assumption is the self-similarity of turbulence in the inertial range. 
In that context, functional models  are useful in situations in which the effect of the unresolved scales is well understood and quantified. Since in practice this is rarely the case for compressible MHD, and in the absence of extensive experimental data for calibration and validation,}
 we proceed with the derivation of a nonlinear structural closure, which is based on the properties of the finite resolution operator, rather than turbulence itself. Thus the MHD turbulence dynamics is not required to obey any strong assumptions, like scale-similarity, existence of an inertial range, energy cascade etc. The resulting closure is closely related to a previously \aprio validated one\cite{Grete2015}, but \REV{includes} additional compressibility effects. \REV{The present paper focuses on the derivation of the new compressible MHD closure, the analytic description of its scope of applicability and energy dissipation properties.
A numerical validation of the closure is performed in} an accompanying work\cite{Grete2016} by \aprio comparison to well-resolved numerical data\REV{, where} it is found to outperform all closures with which it has been compared.

\section{Approximate deconvolution}
As is usual in LES theory, the presented closure has its origins in incompressible hydrodynamics. In particular, it is \REV{a self-consistent extension of } the Yeo-Bedford \REV{(YB)} expansions\cite{Yeo1987, Yeo1993} \REV{as applied to compressible MHD.}
\REV{Closures of this family have} been recently applied to incompressible\cite{Balarac2008, Fabre2011, Balarac2013} and compressible (supersonic) MHD\cite{Grete2015, Vlaykov2015} turbulence with encouraging results. The same method has also been used to model the transport of a passive scalar \cite{Balarac2013}.
Here, we focus on the closure derivation and extend it to include so far unaccounted for compressibility effects.

\REV{For clarity, this section summarizes the original derivation\cite{Yeo1987} as applied to a Gaussian filter kernel and the incompressible MHD SGS terms. 
The Gaussian kernel can be represented by its Fourier transform, i.e. transfer function $\widehat{G}$ given by 
\begin{equation}
\label{eq: gauss ker}
\hat{G}(k)=\textrm{exp}\bra{-\Delta^2 k^2/(4\gamma)},
\end{equation}
with wavenumber $k$ and filter scale $\Delta$. It is infinitely differentiable, which renders it particularly suitable for analytical manipulation. It is also positive, and therefore signature preserving. Thus under its action the SGS counterparts of positive definite quantities like energy are also positive definite\cite{Sagaut2006}.  Furthermore, by setting the width parameter $\gamma=6$, its first and second order moments match those of a
box filter with the same filter scale $\Delta$.

The main idea of the YB expansion is to compute an approximation of the inverse filtering operator based on gradient expansion of the filter kernel $G$.
This amounts to computing an approximation of the inverse Fourier transform of $1/\widehat{G}$.
}The first step is to perform a Taylor expansion of the transfer function and its inverse in terms of the filter scale $\Delta$, i.e.
\begin{align}
	\widehat{G}(\v{k}) = \sum_{n=0}^\infty \frac{(-1)^n}{n!}\bra{\frac{\Delta^2}{4\gamma}\v{k}^2}^n,\\
	\frac{1}{\widehat{G}(\v{k})} = \sum_{n=0}^\infty \frac{1}{n!}\bra{\frac{\Delta^2}{4\gamma}\v{k}^2}^n .
\end{align}
\REV{Applying the expansions to the} test fields $\widehat{f}$ and  $\widehat{\filt{f}}$ respectively, \REV{followed by} an inverse Fourier transformation \REV{yields} infinite series representations of the filter operator and its inverse in terms of gradient operators acting on the test fields,

\begin{align}
	\label{eq: backward expansion}
	\filt{f}= G \ast f = \sum_{n=0}^\infty \frac{1}{n!}\bra{\frac{\Delta^2}{4\gamma}\grad^2}^n f,\\ 
	\label{eq: forward expansion}
	f  =  G^{-1} \ast \filt{f} = \sum_{n=0}^\infty \frac{(-1)^n}{n!}\bra{\frac{\Delta^2}{4\gamma}\grad^2}^n \filt{f}.
\end{align}
They are absolutely convergent and formally accurate at all orders, since the Gaussian kernel is infinitely differentiable and with unbounded support.
In fact, it has been found\cite{Pruett2001} that the series \REV{given in} \cref{eq: backward expansion} converges for all canonical filters, and more generally,  symmetry of the filtering kernel and non-negativity of its transfer function are sufficient conditions for its convergence for a periodic band-limited field $f$. (The last condition  is trivial\REV{ly satisfied} in any numerical simulation.) It has also been suggested\cite{Pruett2001} that qualitatively the convergence rate tends to decrease as the dissipative strength of the filter increases. In the case of the Gaussian filter, the same results hold for the forward expansion \cref{eq: forward expansion}, as it differs from \cref{eq: backward expansion} only by an alternating sign

To proceed note that the unknown components of the SGS stresses and \REV{the} EMF are of the form $\flt{fg}$. Applying \cref{eq: backward expansion} to such an expression results in a series in terms of $(fg)$. As it is absolutely convergent, \cref{eq: forward expansion} \REV{can be applied}
separately to each $f$ and $g$ term of the series.
The result \REV{can be simplified to}

\begin{align}
	\label{eq: YB formulae}
	\flt{fg} =&
	\flt{f}\flt{g}+2a\flt{f}_{,k}\flt{g}_{,k}+
	\frac{1}{2!}\bra{2a}^2 \flt{f}_{,kl}\flt{g}_{,kl}+\\ \nonumber
&	\frac{1}{3!}\bra{2a}^3 \flt{f}_{,klm}\flt{g}_{,klm}+O\bra{a^4\grad^8}\REV{,}	
\end{align}
\REV{as given in eq.~(5.21) of Yeo (1987)\cite{Yeo1987}.}
Here a comma \REV{is used} to represent differentiation with respect to a co-ordinate
and $a = \Delta^2/\bra{4 \gamma}$.
The coefficients in the expansions are given in terms of moments of the transfer function and its inverse.
This relationship comes from the orthogonality of the terms in the Fourier expansion and thus holds for any filter kernel for which the expansion exists.
There is a closed form expression\cite{Carati2001} for the coefficients in \cref{eq: YB formulae} for a
symmetric filter kernel $G$ with infinitely differentiable transfer function \REV{-- they} are given by 
the Taylor coefficients of the function $F(f,g) = G(-i(f+g))/(G(-if) G(-ig))$.
Moreover, since \REV{any} symmetric filter has a real transfer function, only the even coefficients are non-zero.
This symmetry has \REV{a fundamental} impact on the form of the terms in the expansion as well, namely each field
is differentiated at most once with respect to a co-ordinate.

Recall that for $\gamma=6$ the Gaussian and box filter kernels have identical first and second moments. Therefore with this parameter choice \cref{eq: YB formulae} is also valid for a box filter up to second order.
Furthermore, since all moments of a Gaussian function can be expressed in terms of its second order moment, here $(2a)$, it is the only parameter which 
can appear in \cref{eq: YB formulae}.

Applying \cref{eq: YB formulae} to the SGS terms in the incompressible MHD equations is sufficient to  completely close them,
\begin{align}
	\label{eq: YB incompressible sgs}
	\flt{u_i u_j} - \flt{u}_i \flt{u}_j &= 2a\flt{u}_{i,k}
	\flt{u}_{j,k} ,\\ \nonumber
	\flt{B_i B_j}- \bres_i\bres_j &= 2a \bres_{i,k} \bres_{j,k},\\ \nonumber
	\bra{\flt{\v{u}\times\v{B}}-\flt{\v{u}}\times \flt{\v{B}}}_i &= 2a\levi
	\flt{u}_{j,l}\flt{B}_{k,l}.
\end{align}
It should be noted that the resulting closures have been reached by alternative routes in hydrodynamic LES.
The tensor-diffusivity models\cite{Clark1979, Leonard1974, Vreman1996}, for instance, use Taylor expansions of the SGS terms 
with respect to the turbulent fluctuations (e.g.~$\v{u}'=\v{u}-\ures-$) or the entire (unfiltered) fields (e.g.~$\v{u}$). 
These \REV{derivations} however are questionable as they require smoothness of the small scales\cite{Love1980}. 
Another alternative, originally designed for image processing \cite{vanCittert1931}, is given by approximate deconvolution closures\cite{Stolz1999,Stolz1999, Stolz2001, Stolz2001a, Stolz2003, Stolz2005, Stolz2007, Sagaut2009}.
\REV{They are} again based on the truncation of an infinite series to reconstruct the inverse of the filtering operator. However, in this approach the series is not necessarily convergent and truncating at the optimal order is \REV{critical}. The results of both approaches for a Gaussian filter agree
with \cref{eq: YB formulae} up to second order\cite{Sagaut2009}.
The different motivations and derivation are revealed only at higher orders.

\section{Compressible extensions}
\REV{To apply the} presented derivation self-consistently to the \REV{compressible} Reynolds SGS stress and EMF, as defined in \cref{eq:tau_emf_def} the compressibility effects onto the mass-weighted large scale velocity \REV{has to be taken further into account}. The issue can be addressed from several viewpoints.
On the one hand, one can dispense with the mass-weighted filtering operator altogether, and re-substitute $\ffilt{f}\,\rres=\filt{f \rho}$ in the relevant 
SGS terms. This requires that an additional SGS term $\flt{\rho u_i} -\rres \, \flt{u}_i $ is introduced in the continuity equation,	
and that the EMF and the Reynolds SGS stress are re-defined.
The complexity of the Reynolds SGS stress $\tu$ \REV{is} formally increased, as it now contains an unclosed product of three fields, i.e. $\filt{\rho u_i u_j}$.
Nevertheless, the derivation outlined above still holds. Applying \cref{eq: backward expansion,eq: forward expansion} to a general term of third order leads to \REV{(as given by eq.~(5.23) of Yeo (1987)\cite{Yeo1987})}
\begin{align}
	\label{eq: YB triple}\nonumber
	\flt{fgh} =& \flt{f} \flt{g} \flt{h} +
	2a\bra{ \flt{f}_{,k} \flt{g}_{,k} \flt{h}+
		\flt{f}_{,k} \flt{g} \flt{h}_{,k}+
		\flt{f} \flt{g}_{,k} \flt{h}_{,k}}+\\&\nonumber
	\frac{1}{2!}\bra{2a}^2 \left(\flt{f}_{,kl} \flt{g}_{,kl} \flt{h}+
		\flt{f}_{,kl} \flt{g} \flt{h}_{,kl}+
		\flt{f} \flt{g}_{,kl} \flt{h}_{,kl}+\right.\\&\nonumber
	\left.	2\flt{f}_{,k} \flt{g}_{,kl} \flt{h}_{,l}+
		2\flt{f}_{,k} \flt{g}_{,l} \flt{h}_{,kl}+
		2\flt{f}_{,kl} \flt{g}_{,k} \flt{h}_{,l}
	\right)+\\& +O\bra{a^3\grad^6}.
\end{align}
To first order in $a$ this technique leads to
the following results for the primary SGS terms\REV{:}
\begin{align}
	\label{eq: YB incompressible tu}
	\flt{\rho u_i} -\rres \, \flt{u}_i  &= 2a \rres_{,k}\flt{u}_{i,k}\\ \nonumber
	\flt{\rho u_i u_j} - \rres\, \flt{u}_i \flt{u}_j &= 2a \rres \,\flt{u}_{i,k}
	\flt{u}_{j,k} +
	2a \rres_{,k}\bra{\flt{u}_{i,k} \flt{u}_j+\flt{u}_i\flt{u}_{j,k}},\\ \nonumber
	\flt{B_i B_j}- \bres_i\bres_j &= 2a \bres_{i,k} \bres_{j,k},\\ \nonumber
	\bra{\flt{\v{u}\times\v{B}}-\flt{\v{u}}\times \flt{\v{B}}}_i &= 2a\levi
	\flt{u}_{j,l}\flt{B}_{k,l}.
\end{align}
This constitutes a complete closure of the compressible MHD equations (barring pressure considerations).
This approach is applicable for numerical schemes which evolve the velocity field, because only directly filtered fields are present. 
Even though such schemes are not frequently used to address highly compressible problems, such a model has been implemented in compressible hydrodynamics\cite{Bin2007}.

On the other hand, for applications to compressible codes which treat the momentum as a primary quantity, e.g. using finite volume schemes, one needs to take into account the mass-weighted filtering operator.  For a field $f$ it is given by $\ffilt{f}=(G\ast(\rho f))/(G\ast\rho) $.
In the process of directly applying the outlined procedure to this operator, several fundamental challenges are encountered. The main obstacle is that since its filter kernel contains strongly fluctuating contributions (e.g. from the $G\ast \rho$ component), the Taylor expansion of its transfer function is not well-defined. Additionally, 
the existence of the inverse transfer function is not assured over an extended interval in spectral space.

\subsection{Simple compressible extension}
The simplest hypothesis which circumvents the complications outlined above would be to assume that even if the derivation is not valid for compressible MHD, its result still holds, i.e. to apply the map 
 \begin{equation}
 \overline{\v{u}} \rightarrow \tilde{\v{u}}.
 \label{eq: trivial map}
 \end{equation}
to the incompressible closures \cref{eq: YB incompressible sgs}.
This would imply that the compressibility effects are implicitly taken into account by the change of operator.
Qualitatively, this approach could be motivated by invoking the \REV{reduction} of compressibility effects at smaller scales\cite{Erlebacher1992}, but ultimately it is the simplest compressibility extension of \cref{eq: YB incompressible sgs}.
In fact, \REV{a previous \aprio comparison\cite{Grete2015} with data from supersonic numerical simulations showed that this extension} yields consistently higher correlation with the data than the other tested classical closures. \REV{However},
while the results for the SGS stress were consistently high, the EMF closure exhibited a comparatively larger scatter. \REV{This difference can be explained by the self-consistent derivation of compressibility effects which follows.}

\subsection{Primary compressible extension}
The goal is to obtain an expression of a simply filtered field in terms of the corresponding mass-weighted filtered field. Since mass-weighting applies to velocity-related fields, consider in particular $\ures-= \flt{\v{u}\rho}/\rres$. 
Applying \cref{eq: YB formulae} to the right-hand side \REV{leads to}
\begin{equation}
	\label{eq: utilde recurrence basis}
	\ures_i = \flt{u}_i+2a y_{,k} \flt{u}_{i,k}+2a^2
	\bra{y_{,kl}+y_{,k}y_{,l}}\flt{u}_{i,kl}+O\bra{a^3},
\end{equation}
where we denote for brevity the natural logarithm of the resolved density as $y= \ln \rres$. As \eqref{eq: utilde recurrence basis}
represents an absolutely convergent series, under the same conditions as the original expansion \cref{eq: backward expansion}, it can be rearranged to give
\begin{equation}
	\label{eq: ubar recurrence basis}
	\flt{u}_i =\ures_i-2a y_{,k}  \flt{u}_{i,k}-2a^2
	\bra{y_{,kl}+y_{,k}y_{,l}}\flt{u}_{i,kl}-O\bra{a^3}.
\end{equation}
To this we can apply a recurrence technique. To second order in $a$ it gives
\begin{align}
	\label{eq: recurse2}
	\flt{u}_i =&\ures_i-2a y_{,k}\ures_{i,k}-\\\nonumber &
	2a^2\bra{\bra{y_{,kl}-y_{,k}y_{,l}}\ures_{i,kl}-
		2y_{,k}y_{,kl}\ures_{i,l}}-O\bra{a^3}.
\end{align}
This expression, along with \cref{eq: YB formulae,eq: YB triple}, can be applied to the definition of the SGS terms, \cref{eq:tau_emf_def}, to obtain 
\TODO{apply the result to the EMF and stresses, recommend first order truncation and emphasize no extra computational costs}
\begin{align}
\label{eq: nonlin final tu}
\tu[][ij]&=2a\rres \ures_{i,k}\ures_{j,k}+\\ \nonumber & 
2a^2\rres \bra{\ures_{i,kl}\ures_{j,kl}-2y_{,kl}\ures_{i,k}\ures_{j,l}}+O\bra{a^3},\\
\label{eq: nonlin compr emf}
\emf_i &= 2a\levi \bra{\ures_{j,l} \bres_{k,l}-
	y_{,l}\ures_{j,l}\bres_{k}}+\\\nonumber&
2a^2\levi
\left(\ures_{j,lm}\bres_{k,lm}-2\bra{y_{,lm}\ures_{j,l}+y_{,l}\ures_{j,
		lm}}\bres_{k,l}+\right.\\\nonumber&\left.
\bra{2y_{,l}y_{,lm}\ures_{j,m}+\bra{y_{,p}y_{,l}-y_{,pl}}\ures_{j,pl}}
\bres_k\right)+O\bra{a^3}.
\end{align}

As the Maxwell SGS stress is not directly affected by density variations, its closure is identical to the one from \cref{eq: YB incompressible sgs}.
Remarkably, to first order the compressibility effects on the Reynolds SGS stress are implicitly accounted for by the mass-weighted filtering itself.
This is a consequence of the symmetry of the Reynolds SGS stress tensor ($\tu[][ij]= \tu[][ji]$). \REV{Explicit d}ensity variations appear here only at second order and as
second order logarithmic derivatives. Therefore
only very strong compressibility cannot be accounted for by the simple compressibility extension implied by \cref{eq: trivial map}.
In contrast, in the EMF closure density variations appear already at first order, 
and at second order they are much more extensive than for $\tu$. 
This explains the different levels of success of the simple compressibility extension\cite{Grete2015}\REV{--} terms which account for compressibility effects \REV{are missing } in the EMF \REV{closure but not in the Reynolds SGS stress one}.

We note that combining the recurrence relation \cref{eq: recurse2} with expansions of the type of \cref{eq: YB formulae,eq: YB triple} allows \REV{the construction of} self-consistent closures for an SGS term of any type to any order. The SGS kinetic and magnetic energies for instance are given trivially as half the traces of the Reynolds or Maxwell SGS stress tensors, respectively. If we were to construct the 
SGS cross-helicity $W_\sgs = \filt{\v{u} \cdot \v{B}} - \ures-\cdot\bres-$, \REV{e.g.} to gauge the correlation between kinetic and magnetic SGS effects, its closure to first order would
be given by
	\begin{equation}
		\label{eq: nonlin compr crhel}
		W_\sgs= 2 a\bra{\ures_{i,j}\bres_{i,j} - \ures_{i,j}y_{,j} \bres_{i}}+ O(a^2).
	\end{equation}
	
Retaining terms to first order in $a$ is expected to provide sufficient SGS information, as suggested by the previously reported results\cite{Grete2015,Balarac2008, Fabre2011, Balarac2013}.
Furthermore, the computational overhead of including such closures in an LES is minimal, as they can contain at most first order derivatives in large scale primary fields. 

\subsection{Extension for the SGS derivatives}
Direct comparison of the outlined closures with the corresponding SGS terms
\REV{based on numerical data} reveals directly the probity of the method\cite{Grete2016}. However, for \apost application of the closures in LES simulations a further compressible effect needs to be considered.

The simple filtering operator is a convolution and as such commutes with differentiation, however the mass-weighted filtering operator does not.
This is critical since the SGS stress and EMF enter the evolution equations under a gradient.
For the purposes of this section, let $\widehat{f}$ denote the closure of an SGS term $f$ incorporating mass-weighted filtering. Then propagating the commutator between mass-weighted filtering and differentiation through the closure calculations \REV{above} yields the following additional contributions to the momentum and induction equations
\begin{align}
\label{eq: diff commutator}
\widehat{\partial_i \tu[][ij]}  -   \partial_i \widehat{\tu[][ij]} =&
2a \rres\bra{\ures_i\ures_{j,l} + \ures_j\ures_{i,l}}y_{,il},\\ \nonumber
\bra{  \widehat{\curl \emf-} - \curl \widehat{\emf-}}_i =& 2 a \levi \levi[klm] \ures_{l,p}  \bres_m y_{,jp}.
\end{align}

These expressions show the difference between applying the closure procedure to the derivatives of the SGS terms and 
taking derivatives of the respective closures. The additional corrections are expected to be important \REV{primarily} for very strong density variations, as they contain second derivatives in the logarithmic density. This can be also seen by comparing the expressions above to the ones obtained by differentiating \cref{eq: YB incompressible tu}. Furthermore, they are of leading order (in $a$) for the derivatives of both SGS terms and these are precisely the quantities which enter the LES evolution equations and affect the large scale dynamics.

Combining the two compressibility effects leads to significant cancellation of the first order terms in the EMF closure with a final result given by
\begin{equation}
 \bra{\widehat{\curl \emf-}}_i = 2 a \levi \levi[klm] \bra{\bra{\ures_{l} \bres_{m}}_{,j} -\bra{\ures_{l,p}  \bres_m}_{,j} y_{,p}}.
 \TODO{double check}
\end{equation}
For the Reynolds SGS stress, the final closure can be given as 
\begin{equation}
\widehat{\partial_{i}\tu[][ij]}=2a\bra{\rres \ures_{i,k}\ures_{j,k}}_{,i} + 2a \rres\bra{\ures_i\ures_{j,l} + \ures_j\ures_{i,l}}y_{,il}.
\end{equation}
Once again, the SGS Maxwell stress closure is trivially derived from \REV{\cref{eq: YB incompressible sgs}}, as it does not contain any mass-weighted large scale fields.

The effects of the two types of compressibility corrections can be identified by different types of \aprio testing.\
In fact, the validity of the compressible closures \REV{were} tested \aprio against a range of data from sub- to hypersonic turbulence simulations and benchmarked against a wide range of alternative closures\cite{Grete2016} with very positive results. In particular, we investigate their performance with respect to the resolved energy and cross-helicity dynamics (cf. \cref{eq: totenres,eq: crhelres}). The primary compressible closures \cref{eq: nonlin final tu,eq: nonlin compr emf} are validated by considering their effect on the spatially local (in the Eulerian sense) dynamics, i.e. on terms of the form $(\tu \cdot \grad) \cdot\ures$ and $\emf- \cdot \grad \times \bres-$. These terms are usually identified with contributions to the resolved energy or cross-helicity cascades.  The impact of these closures on the overall resolved energy or cross-helicity dynamics, e.g. $\ures- \cdot (\grad \cdot\tu)$ and $\bres- \cdot \grad \times \emf-$, is also tested. 
While the impact of the differentiation commutators \cref{eq: diff commutator} is best tested directly in \apost application, by comparing the results of the local and non-local \aprio tests, we give an indication of the parameter regime where these extensions can be important.

\section{Scope of applicability}
The closure described above has been derived without any strong assumptions about the flow or the magnetic field. 
Thus their application is not limited to turbulence simulations, but can be applied in principle to any MHD \REV{simulation in which the small scales are not sufficiently well-resolved}.
Nevertheless, several limitations need to be kept in mind. 

Firstly, we have implicitly assumed that the filter kernel is homogeneous and isotropic and
has a constant filter scale. This translates to numerical schemes with a regular grid. 
Furthermore, no boundary terms have been taken into account, which is consistent with periodic domains. Extensions of SGS closures
to non-regular grids and non-periodic conditions have been studied in incompressible hydrodynamics\cite{Sagaut2006}. However, their application
to the current closure is beyond the scope of this article.

Secondly, the described closures are derived from the analytical form of a filter kernel.
As the effective kernel of an LES for a particular numerical scheme is a combination of various discretizations, e.g.~grid spacing, time-stepping, differential approximations, quadrature, flux limiting, divergence cleaning (for the magnetic field), shock capturing, etc.,
its exact analytical form is rarely available. 
Additional errors stem from the truncation of the infinite series \REV{\cref{eq: YB formulae,eq: utilde recurrence basis}}, i.e. higher order closures are in principle more accurate. Depending on the convergence rate of the expansions for a particular filter, this error may also need to be considered.
Conversely, due to the nonlinear combination of gradient fields, higher order closures are more prone to numerical instabilities\cite{Geurts1997, Vreman1996}.

Finally, in LES applications the SGS terms are based upon information contained in resolved fields, which resides above the Nyquist scale, i.e.~the grid resolution. This can be represented by decomposing the effective filter kernel into a spectral kernel at the Nyquist scale and a remainder. The spectral kernel renders the inverse transfer function of the effective filter ill-defined. In order to circumvent this, a two-step procedure can be applied. First, the derivation above should be applied to the component of the effective filtering operator with a formally well-defined inverse.
The spectral filter can then applied to the resulting equations.

To allow for the mentioned inaccuracies and numerical instabilities additional renormalization may be applied to the final closures.
Parametric renormalization may also be applied to the results of a closure for a well-behaved filter, as outlined above, in order to boost its dissipative effect or render it suitable for a selection of numerical schemes.
The renormalization can come in the form of constant coefficients or variable fields. Both practices are common in LES. Most canonical SGS closures include a constant coefficient whose value is calibrated \REV{dynamically or } against experimental data. Allowing for distinct coefficients for the different additive terms in the proposed closures and calibrating them against a particular dataset, may be used as a guide for the relative importance of the different terms in the respective flow. 
With respect to spatially varying modulation, the SGS energy for instance can be used to renormalize the strength of the SGS effects in a hydrodynamic LES with a related closure\cite{Schmidt2011, Woodward2006}. This technique naturally requires an additional closure for the SGS energy -- a common situation in hydrodynamics\cite{Bardina1983, Speziale1988, Schmidt2006, Schmidt2011, Carati2001, Sagaut2009, Iapichino2009}, where different closures are frequently combined in order to alleviate their respective shortcomings. 
Both types of renormalization outlined above are applied and \aprio tested\cite{Grete2016} for the proposed closures, however it is found that  neither is particularly necessary or beneficial.

\section{Energy and cross-helicity dissipation properties}
One of the main functions of SGS closures is to correct for the transfer of energy across the resolution scale.
Therefore we proceed with an analysis of the dissipation properties of the proposed closures.
\REV{In particular, we consider the local dissipation of the resolved kinetic energy, magnetic energy and cross-helicity given respectively by 
	\begin{align}
	\Sigma^\kin=-\tau[][ij] \S[][ij], \;\;\;\Sigma^\mag=- \emf-\cdot \jres-\;\; \textrm{ and }\\
	\chcasc = -\frac{\tau[][ij]}{\rres} \bra{\M[][ij]-\bres_j y_{,i}} - \emf- \cdot \wres,
	\end{align}
	with the usual definitions of the resolved rate-of-strain $\S[][ij]=1/2\bra{\ures_{i,j}+\ures_{j,i}}$, vorticity $(\wres-)_k= (\nabla \times \ures-)_k$, current  $(\jres-)_k= (\nabla \times \bres-)_k$ and magnetic rate-of-strain	$\M[][ij]=1/2\bra{\bres_{i,j}+\bres_{j,i}}$.
The signs of the $\Sigma$ fields are chosen such that positive values correspond to a down-scale transfer, i.e. dissipation.
	
	We consider each dissipation term in turn. The kinetic energy dissipation can be further decomposed according to \cref{eq:tau_emf_def} into $\Sigma^\kin= \Sigma^\kin_{\tu}+\Sigma^\kin_{\tb}+\Sigma^\kin_{\tb[][kk]}$. The contribution from the Reynolds SGS stress is given by $\Sigma^\kin_{\tu} = -\tu[][ij] \S[][ij]$. The results here will be the same as in the hydrodynamic limit.}
As a basis for comparison, consider the classical \REV{incompressible} eddy-viscosity (EV) family of closures\cite{Chernyshov2007}, which take the form $\tu=-\nu_\turb\S$ \REV{with $\Tr(\S)\equiv0$} for some (usually non-negative) turbulent viscosity $\nu_\turb$. For it \REV{$\Sigma^\kin_{\tu}$} takes the form
\begin{equation}
\Sigma^\kin_\mathrm{EV}=\nu_\turb\REV{\Tr(\S[2]),}
\end{equation}
\REV{where $\S[n]$ represents a tensor product, e.g.~ $(\S[2])_{ij}=\S[][ik]\S[][kj]$}. As \REV{$\Tr(\S[2])$} is always non-negative, this closure can transfer energy across the resolution scale only in one direction,
depending on the sign of $\nu_\turb$, e.g. from resolved to subgrid scales for $\nu_\turb>0$. \REV{This model can provide energy backscatter only in the compressible regime via an additional (not self-consistent) closure for the SGS kinetic energy and even then only from regions where $\Tr(\S)>0$. This can be seen to be problematic since the presence of strong energy cascades in both directions is a key characteristic of MHD turbulence\cite{Pouquet1976,Muller2012}, which differentiates it from the hydrodynamic case.}

\REV{In contrast, the proposed} closure for the Reynolds SGS stress $\tu$ can be written as
\begin{align}
 \tu[][ij] 
	=& 2 a \rres\bra{\S[][ik]\S[][\REV{jk}] + \wres_{ik} \wres_{\REV{jk}}+ \S[][ik]\wres_{\REV{jk}}+ \wres_{ik}\S[][\REV{jk}]},
\end{align}
\REV{with vorticity tensor $\wres_{ij}= -1/2\levi (\wres-)_k$.} Substituting this in \REV{$\Sigma^\kin_{\tu}$} leads to
\begin{align}
\label{eq: eps tu_k 1}
\Sigma^\kin_{\tu} =&\REV{-2a\rres\bra{ \Tr(\S[3])+ \frac{1}{4}\wres-^2\Tr(\S)-\frac{1}{4}\wres-^\mathrm{T}\!\cdot\S \cdot \wres-}.}
\end{align}
\REV{The first term is reminiscent of the eddy-viscosity expression, as it depends only on the strain tensor. However, there are two qualitative differences stemming
	from the fact that this term is cubic in $\S$. Firstly, the larger power leads to stronger sensitivity to the resolved rate-of-strain. Secondly, and perhaps more importantly, this term has indefinite signature, which allows for bi-directional energy cascade. Because of it totally compressive rate-of-strain leads to dissipation while expansion leads to back-scatter of kinetic energy.
	
	The proposed model includes a further effect, associated with the last two terms in \cref{eq: eps tu_k 1}, namely vortex stretching. This is the compressible analogue of the incompressible vortex stretching effect encoded in the last term. Geometrically, the combination of the two terms represent the interaction of the vorticity vector with the strain lying in a plane orthogonal to it. As intuition suggests, if a simple vortex tube is compressed perpendicular to its axis, its radius decreases and bigger proportion of its kinetic energy is associated with smaller scales, i.e. this leads to dissipation. Conversely, stretching a vortex, shifts its associated energy to larger scales and the result is back-scatter.
	
	Next, consider the contribution of the Maxwell SGS stress to the kinetic energy flux given by $\Sigma^\kin_{\tb}=\tb[][ij]\S[][ij]$. The proposed closure can be written as}
\begin{align}
\tb[][\REV{ij}]=
& 2a\left(\M[][ik] \M[][\REV{jk}]+\jres_{ik}\jres_{\REV{jk}}+ \M[][ik]\jres_{\REV{jk}}+ \jres_{ik}\M[][\REV{jk}]\right),
\end{align}
\REV{with current tensor $\jres_{ij}= -1/2\levi (\jres-)_k$.} Its contribution to the kinetic energy dissipation is given by

\begin{align}
\nonumber
\Sigma^\kin_{\tb}=
2a &\REV{\left( \Tr(\M\S\M)+ 2\Tr(\M \S \jres) \right. }\\
		&\REV{ \left.+\frac{1}{4} \jres-^2 \Tr(\S)- \frac{1}{4} {\jres-}^\mathrm{T} \! \cdot \S \cdot \jres- \right).}
 \label{eq: kin diss taub}
\end{align}
This expression is similar to the contribution of the Reynolds SGS stress\REV{.  Note however, that the entire Maxwell SGS stress
	works in the opposite direction to the Reynolds SGS stress (because of the different overall sign). The first term represents the interaction between the magnetic and kinetic rates-of-strain. Here compression (i.e. negative eigenvalues of $\S$) leads to back-scatter, while stretching leads to dissipation. Furthermore, alignment of the eigenvectors of $\S$ and $\M$ maximizes the effect of this term.  The second term is associated with the amplification of magnitudes of the rates-of-strain, i.e. $\Tr(\S[2])$ and $\Tr(\M[2])$. It implies that the processes which enhance kinetic and magnetic shearing simultaneously dissipate kinetic energy. The last two terms are the counterpart of the vorticity terms \cref{eq: eps tu_k 1} -- they are associated with current deformation analogous to the vortex stretching effect. 
	They imply that currents perpendicular to compressive flows lead to backscatter and ones perpendicular to expanding flows -- to dissipation. Currents flowing along a compressive or stretching directions have no effect on the SGS energy.
	
	The final component of the kinetic energy flux comes from the SGS magnetic pressure
	\begin{equation}
	\Sigma^\kin_{\tb[][kk]}=-\half\tb[][kk] \Tr(\S)=
	-2a \Tr(\S)\bra{\frac{\Tr(\M[2])}{2}+\frac{1}{4} \jres-^2}.
	\label{eq: kin diss taub_kk}
	\end{equation}
It reduces the Maxwell SGS stress effects associated with the overall dilatation rate. It introduces purely compressible effects, as in the incompressible limit $\Tr(\S)=0$. The isotropic current component ($\propto \Tr(\S)\jres-^2$) cancels exactly the contribution from $\Sigma_{\tb}^\kin$. This re-introduces the possibility of dissipation due to compression along the current direction and emphasizes the importance of providing a closure for the total SGS pressure. Moreover, it enhances the closure's overall sensitivity to the relative orientation of the current and the kinetic rate of strain. The magnetic shear term is associated with the growth of $\Tr(\M[2])$ due to overall compression.}

Finally, consider the transfer of magnetic energy across the filter scale. \REV{The analytic form of $\Sigma^\mag$ shows that }there is backscatter, or dynamo-like \REV{effect}, when the electromotive force is aligned with the large-scale currents and dissipation into unresolved energy in cases of anti-alignment. 
 Decomposing the proposed closure into symmetric and anti-symmetric gradients of the resolved fields and substituting into the expression for $\Sigma^\mag$, leads to the following expression
\begin{align}
\label{eq: mag en diss}
\Sigma^\mag =&\REV{2 a \left(2\Tr( \M \S \jres) + \half \jres-^\mathrm{T} \cdot \S \cdot \jres- - \half \jres-^2 \Tr(\S)\right.}\\\nonumber
&\REV{\;\;\;\;\;\;-\half \wres-^\mathrm{T}\! \cdot \M \cdot \jres-}\\\nonumber
&\REV{\;\;\;\;\;\; +\bra{\bres- \times \jres-}^\mathrm{T}\!\!\cdot\S \cdot \nabla y}\\\nonumber
&\REV{\;\;\;\;\;\;+\left.\half \bra{\wres-\cdot \bres-}\bra{\jres-\cdot \nabla y}
-\half\bra{\wres-\cdot \jres-}\bra{\bres- \cdot \nabla y}  \right)   .}
\end{align}
Due to the nonlinear coupling between kinetic and magnetic structures in this closure, \REV{these terms involve a large plethora of effects. 
		
Here, like in the kinetic energy case, the relative alignment of the resolved gradients, i.e. the local inhomogeneity and anisotropy, play a vital role in determining the magnetic energy flux. The first four terms are associated with evolution of the total current $\jres-^2$. The first, shearing term is already familiar from \cref{eq: kin diss taub} and has the same effect on the magnetic energy as on the kinetic one. The next two terms can be identified as anomalous (anisotropic) resistivity. They are also found in \cref{eq: kin diss taub}, but with opposite signs and half the amplitude. This identifies an SGS channel for transfer between resolved kinetic and magnetic energy, i.e. half of the dissipated resolved magnetic energy is backscattered into resolved kinetic energy and vice versa, kinetic energy dissipation leads to enhanced turbulence, which in turn causes a dynamo-like increase of resolved magnetic energy. The fourth term is specific to the magnetic energy budget. It is also associated with the enstrophy evolution due to the Lorentz force and connects the relative orientation of vorticity and current with the principal axes of $\M$. For instance, along a magnetically compressive direction it leads to dissipation, if the vorticity and the current are parallel, and backscatter, if they are anti-parallel.

All considerations made so far apply equally to the simple and primary compressible extensions, as well as in the incompressible limit (allowing for $\Tr(\S)=0$). The final three terms of the magnetic energy dissipation \cref{eq: mag en diss} contain the explicit effect of the primary compressible extension. They have a strong impact primarily in regions of very strong density gradients, e.g.~the neighborhood of shocks, due to the logarithmic density derivative. Formally, they are also strongly anisotropic and can be seen to be related to dynamo-like effects. For instance $\bres- \times \jres-$ is the complement of the current helicity $\bres-\cdot \jres-$, which can be associated with the $\alpha$-dynamo, while $\wres- \cdot \jres-$ is related to the cross-helicity dynamo\cite{Yokoi2013}.	

The effect of the primary compressible extension becomes more evident when considering the SGS effects on the cross-helicity evolution.
	For completeness we give the exact expressions for the local contributions of the total SGS Maxwell Stress $ \Sigma^W_{\tb[][\mathrm{tot}]}=\Sigma^W_{\tb}+\Sigma^W_{\tb[][kk]}$, the SGS Reynolds stress $\Sigma^W_{\tu[][]}$ and the EMF $\Sigma^W_{\emf-}$, defined analogously to their energy counterparts, to the resolved cross-helicity:
\begin{align}
 \Sigma^W_{\tb[][\mathrm{tot}]} = -\frac{2a}{\rres} &\left( \bra{\bres-^\mathrm{T} \cdot \M[2] \cdot \nabla y}-\Tr(\M[3])-\right. \\ \nonumber 
		&\half \Tr(\M[2]) \bra{\bres- \cdot \nabla y}
		+\jres-^\textrm{T}\! \cdot \M \cdot \jres- - \\\nonumber
		&\bra{\bres-^\mathrm{T} \cdot \M}\cdot \bra{\jres- \times \nabla y}- \bra{\jres- \times \bres-}^\mathrm{T}\cdot\bra{\M \cdot \nabla y}\\\nonumber
		&\left. -\bra{\jres- \cdot \bres-}{\jres- \cdot \nabla y}  \right),
\end{align}

 \begin{align}
  \Sigma^W_{\tu[][]} = 2a &\left(-2\Tr(\S \M \wres) - \frac{1}{4} \bra{\wres-\cdot \bres-} \bra{\wres- \cdot \nabla y}   \right. \\ \nonumber 
&+\frac{1}{4} \wres-^\mathrm{T} \cdot \M \cdot \wres-  + \frac{1}{4}\wres-^2 \bra{\bres- \cdot \nabla y} \\\nonumber
&+ \half \bra{\bres- \times \wres-}^\mathrm{T}\cdot \bra {\S \cdot \nabla y} - \Tr(\S\M\S) \\\nonumber 
&\left.-\half \bra{\bres-^\mathrm{T} \cdot \S} \cdot \bra{\wres- \times \nabla y} + \bres-^\mathrm{T} \cdot \S[2] \cdot \nabla y \right),
 \end{align}
 
 \begin{align}
 \Sigma^W_{\emf-} = 2a & \left(2 \Tr(\S \M \wres) +\half \bra{\wres- \cdot \bres-} \bra{\wres- \cdot \nabla y}  \right . \\\nonumber
 &  -\half \wres-^\mathrm{T} \cdot \M \cdot \wres-  - \half \wres-^2 \bra{\bres- \cdot \nabla y}\\\nonumber
  &-\bra{\bres- \times \wres-}^\mathrm{T} \cdot \bra {\S \cdot \nabla y}+ \half \wres-^\mathrm{T} \cdot \S \cdot \jres-\\\nonumber
&\left.  -\half \bra{\jres- \cdot \wres-}\Tr(\S)\right).
 \end{align}
While these expressions contain a large variety of terms, the key point is that there is a strong interplay between  Reynolds SGS stress and the EMF contributions, i.e. the terms in $\Sigma^W_{\tu}$ and $\Sigma^W_{\emf-}$. For instance, the cancellation of the $\Tr(\S \M \wres)$ term points to an interaction between the resolved and turbulent fields 
which preserves the large scale topology characterized by $W$. 

Another example is given by the $\nabla y$-terms in $\Sigma^W_{\tu}$ and $\Sigma^W_{\emf-}$. In $\Sigma^W_{\tu}$ they come from the intrinsic compressibility effect described by $\tu[][ij]\bres_j y_{,i}/\rres $, i.e.~the interaction between velocity fluctuations, density gradients and a large scale magnetic field. The corresponding $\nabla y$-terms in $\Sigma^W_{\emf-}$ are specific to the primary compressible extension. The analogous form of the two sets of terms shows that the primary extension naturally restores the symmetry between kinetic and magnetic turbulent contributions to the effects of compressibility on $W$.  As the resolved cross-helicity plays a role in the non-local transfer between kinetic and magnetic energies and affects the rate of energy decay, it is clearly important to treat it with as much care as the resolved energy itself.
}

\section{Conclusion}
The high computational cost of 3-dimensional direct numerical MHD simulations poses severe limitations to our understanding
of  astrophysical and terrestrial phenomena involving strongly turbulent magnetized fluids.
Large-eddy simulations can alleviate this issue by explicitly considering the effects of limited resolution.
In this work, we presented the derivation and properties of a nonlinear structural 
closure of the compressible MHD LES equations.
It is based on a series expansion\cite{Yeo1987} of the finite resolution operator, a convolution with a low-pass filter kernel, and 
careful consideration of the impact of the operator on the compressible dynamics.
As the derivation needs no assumptions on the nature of the flow, the closures can be applied to a wide variety of MHD problems,
as long as they can be described on a regular grid under periodic boundary conditions. In particular, no assumptions were invoked on the level of compressibility,
on the structure, dynamics, or even presence of turbulence and magnetic fields.
Thus the closures are suitable for both statistically stationary and developing disordered velocity and magnetic field configurations, from
the sub- to the hyper-sonic and -Alfvenic regime. Only an isothermal equation of state was considered. However, the formalism can be extended to 
incorporate thermal variations, as well as additional evolution equations, e.g. for the SGS energy or for passive scalar transport.

Although the closures for the MHD SGS terms are derived self-consistently, the
information gap below the Nyquist frequency as well as the complicated nature of realistic LES filters leaves room for additional re-normalization 
or re-calibration of the proposed closures and for combinations with additional closures.
In fact a simple renormalized version of the closure has already been validated\cite{Grete2015} in \aprio comparison. Here, through a self-consistent derivation of the compressibility effects due to a mass-weighted filter, some of the results of this comparison \REV{are} clarified.
An analysis of the energy dissipation properties of the \REV{simple} compressible closure demonstrates that it can already accommodate sophisticated energy transfers between resolved and unresolved kinetic and magnetic energy budgets. It emphasizes the dependence of the transfer on local geometry, e.g.  anisotropy, and topology, e.g the interplay between vortical and shearing magnetic and kinetic structures of different types.
\REV{Furthermore,  it allows for imperfect transfer between the resolved kinetic and magnetic energy mediated by the subgrid scales.  The additional effects of the self-consistent, primary closure are revealed through the resolved magnetic energy dissipation, where it plays a role in regions of strong compressibility. Moreover, it restores the symmetry between kinetic and magnetic contributions to the cross-helicity dissipation, and thus plays a vital role in the evolution of the large-scale fields' topology.}
Thus presented, the closure is ready to be bench-marked against currently used compressible MHD closures and
to have its properties validated against numerical and experimental turbulence data.
The results of such a comparison with a wide selection of available SGS closures against a suite of simulation data of homogeneous and isotropic turbulence ranging from the sub- to the hyper-sonic regime are presented in an accompanying article\cite{Grete2016}.

\begin{acknowledgments}
	P.G. acknowledges financial support by the 
	\textit{International Max Planck Research School for Solar System 
		Science at the University of G\"ottingen}.
	D.V. acknowledges research funding by the 
	\textit{Deutsche Forschungsgemeinschaft (DFG)} under grant
	\textit{SFB 963/1, project A15} and the \textit{Max Planck Institute for Dynamics and Self-Organization}. DRGS thanks for funding through Fondecyt regular (project code 1161247) and through the ''Concurso Proyectos Internacionales de Investigaci\'on, Convocatoria 2015'' (project code PII20150171). This project is supported by the \textit{North-German Supercomputing Alliance}	 under grant \textit{nip00037}.
\end{acknowledgments}

\section*{References}
\normalem
\bibliography{bibliography}
\end{document}